\pdfoutput=1 
\documentclass{JINST}
\usepackage{upgreek}
\usepackage{cite}

\title{The system for delivery of IR laser radiaton into high vacuum}

\author{E.V. Abakumova$^{a}$, 
M.N. Achasov$^{a,b,}$\thanks{Corresponding author.},
A.A. Krasnov$^{a,b}$, N.Yu. Muchnoi$^{a,b}$ ~ and
E.E. Pyata$^{a}$\\
\llap{$^a$}Budker Institute of Nuclear Physics, Siberian Branch of the Russian
 Academy of Science,\\
 11 Lavrentyev, Novosibirsk 630090, Russia\\
\llap{$^b$}Novosibirsk State University,\\
 Novosibirsk 630090, Russia\\
E-mail: \email{achasov@inp.nsk.su}}

\abstract{The system for insertion of a laser beam into the vacuum chamber of
high-energy storage ring is described. The main part of the system is the 
high-vacuum viewport for the IR radiation, based on ZnSe or GaAs crystals. 
The design of the viewports is presented.}


\begin{document}

\section{Introduction}\label{sec:xxx}
 
In experiments at $e^+e^-$ colliders the high accuracy determination of the
beam energy is crucial for lot of studies. The beam energy below 2 GeV can be 
precisely measured by the calorimetric method based on Compton backscattering
of monochromatic $CO_2$($CO$) laser radiation on the beam (CBS method)
\cite{obzor}.

In collider experiments, the CBS method was applied at VEPP-4M \cite{vepp4},
the $\tau-charm$ factory BEPC-II \cite{bems}  and at VEPP-2000 \cite{ems2000}.
For insertion of the laser beam into the vacuum chambers of these colliders
the laser-to-vacuum insertion system was developed. Here we report the design
of the system.

\section{Laser-to-vacuum insertion system overview}

\begin{figure}[tbp]
\centering
\includegraphics[width=1.0\textwidth]{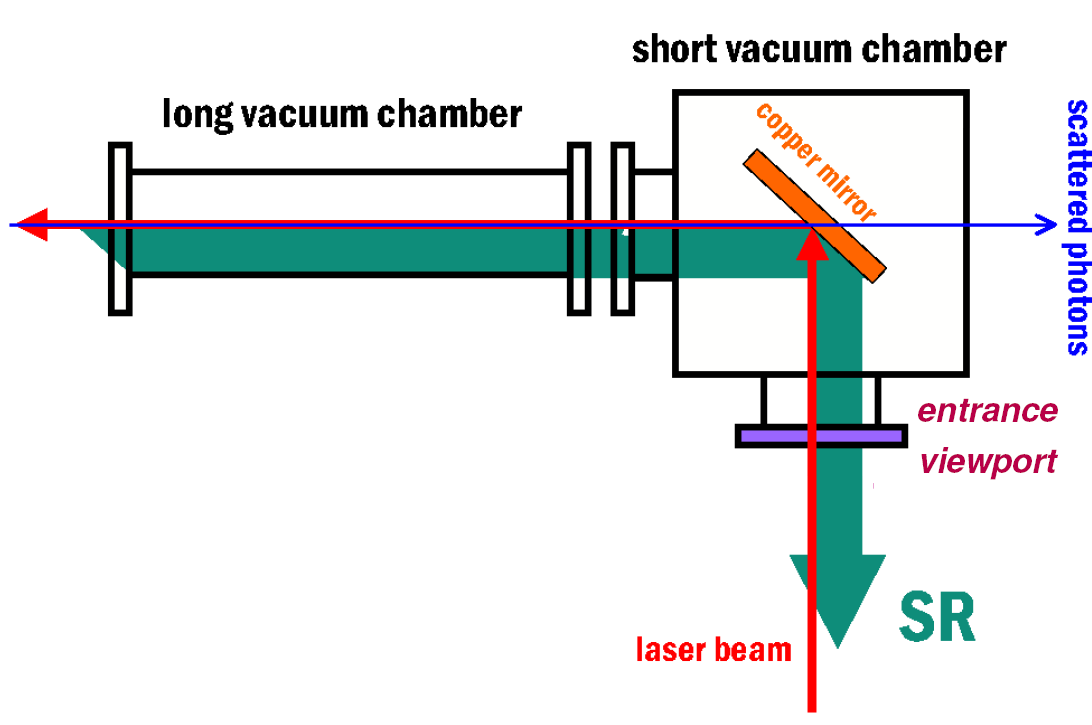}
\caption{Simplified schematic of the laser-to-vacuum insertion assembly.}
\label{cxe-bakyy}
\end{figure}

The delivery of the laser beam into the collider vacuum chamber is performed
using the laser-to-vacuum insertion system. The system is the special
stainless steel vacuum chamber with an entrance viewport and water cooled 
copper mirror (figure~\ref{cxe-bakyy}). The system provides 
extra-high vacuum, i.e. pressure of residual gas inside the chamber is less 
than $5\times 10^{-10}$ Torr. The viewport transfers IR laser light into the 
vacuum and visible synchrotron radiation (SR) light from the vacuum. The output
light can be used to monitor the beam position. The copper mirror in the
vacuum chamber reflects the light  through an angle of $90^\circ$. 
After back-scattering, high energy photons return back to the mirror, pass 
through it and leave the  vacuum chamber. 

There are two types of viewports based on GaAs mono-crystal and ZnSe
polycristal plates. Both viewports are manufactured using similar technology 
and provide:
\begin{enumerate}
\item
baking out the vacuum system up to 250$^\circ$C,
\item
extra-high vacuum,
\item
 transmission spectrum from 0.9 up to 18 $\upmu$m (GaAs viewport) and from 0.45
 to 20 $\upmu$m (ZnSe viewport).
\end{enumerate}

\section{GaAs and ZnSe viewports}

\begin{figure}[tbp]
\centering
\includegraphics[width=1.0\textwidth]{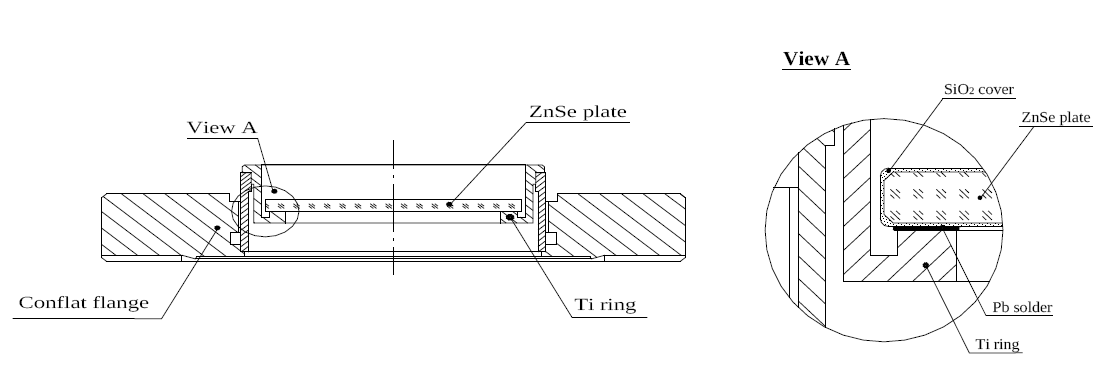}
\caption{The design of ZnSe viewport.}
\label{design}
\end{figure}

The viewport design is shown in figure~\ref{design}. It includes a
304 L steel DN63 conflat flange and a GaAs or ZnSe crystal plate with
diameter of 50.8 mm and thickness of 3 mm or 8 mm respectively. 
In order to compensate mechanically for the difference of the GaAs or ZnSe and 
stainless steel thermal expansion coefficients, the  plate is brazed with 
pure soft lead to a titanium ring, which has been brazed with AgCu alloy
to the stainless steel ring. The stainless steel ring is welded to the flange. 
The flat design of the viewport (thikness is less or equal to 25 mm) allows
to use the viewports in the limited space of physical equipment.
       
To avoid evaporation of  GaAs or ZnSe substance from the plates during 
brazing, they are covered with a 0.6 $\mu$m $SiO_2$ film using gas-phase
deposition \cite{nenash}. This film provide good adhesion of crystal plate with
lead solder and allow to obtain reliable junction.

\begin{figure}[tbp]
\centering
\includegraphics[width=1.0\textwidth]{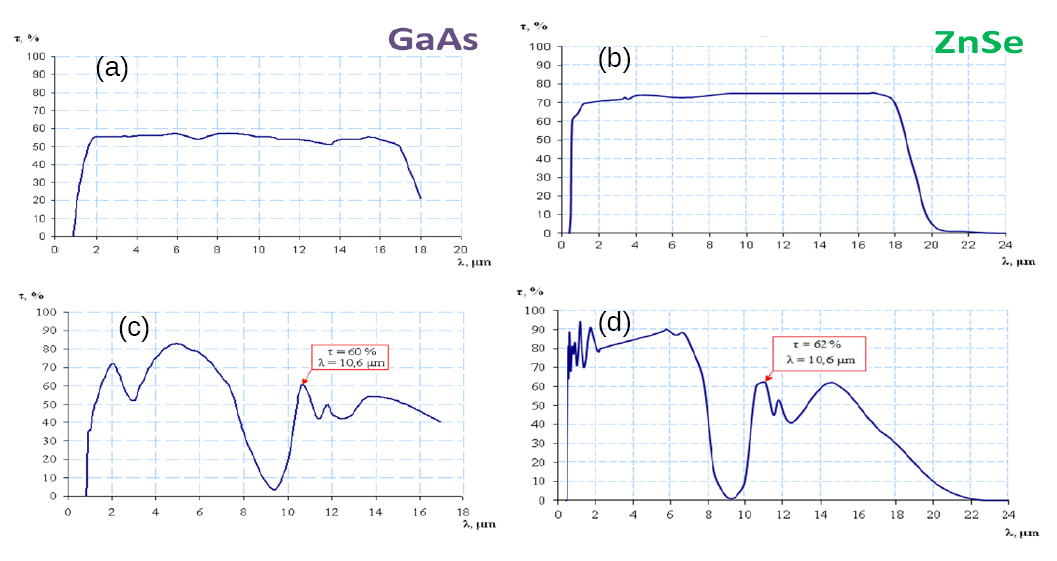}
\caption{Transmittance spectra of uncoated GaAs and ZnSe crystals (a) and (b)
and coated ones (d) and (c) by $0.6\mu m$ SiO$_2$ layer.}
\label{transpec}
\end{figure}

The transmission spectra of the plate before and after covering are shown 
in figure~\ref{transpec}. The transmission of GaAs plate increases from 55
to 60 \% at the $CO_2$ laser wavelength $\lambda=10.6 \mu$m and from 20 to 
35\% at $\lambda=1 \mu$m. In case of ZnSe plate the transmisson at
$\lambda=10.6 \mu$m decrease from 75 to 62 \%, but it turned out to be
comparable with transmittance of GaAs plate. The advantage of ZnSe viewport 
is that it is transparent for the visible part of SR light. This makes the 
beam position monitoring more convenient.

\begin{figure}[tbp]
\centering
\includegraphics[width=1.0\textwidth]{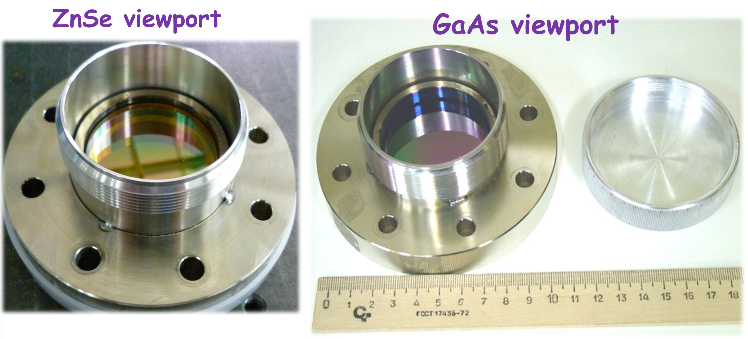}
\caption{Products GaAs and ZnSe viewports.}
\label{photo}
\end{figure}

The viewports were tested at vacuum stand with pressure less than $10^{-8}$
Torr. The tests have included several bakings out at 250$^\circ$C for 8 hours.
The temperature was raised  and lowered at the rate 80$^\circ$C per
hour. After baking the air-tightness tests of viewports with sensitivity
better than 
$10^{-10}$ mbar$\cdot$litre/s were performed. These studies have demonstrated
that after numerous bakings the viewports have good vacuum properties and 
can be used at storage rings. Figure~\ref{photo} shows the ZnSe an
GaAs viewports ready for usage.

\section{Copper mirror}

\begin{figure}[tbp]
\centering
\includegraphics[width=1.0\textwidth]{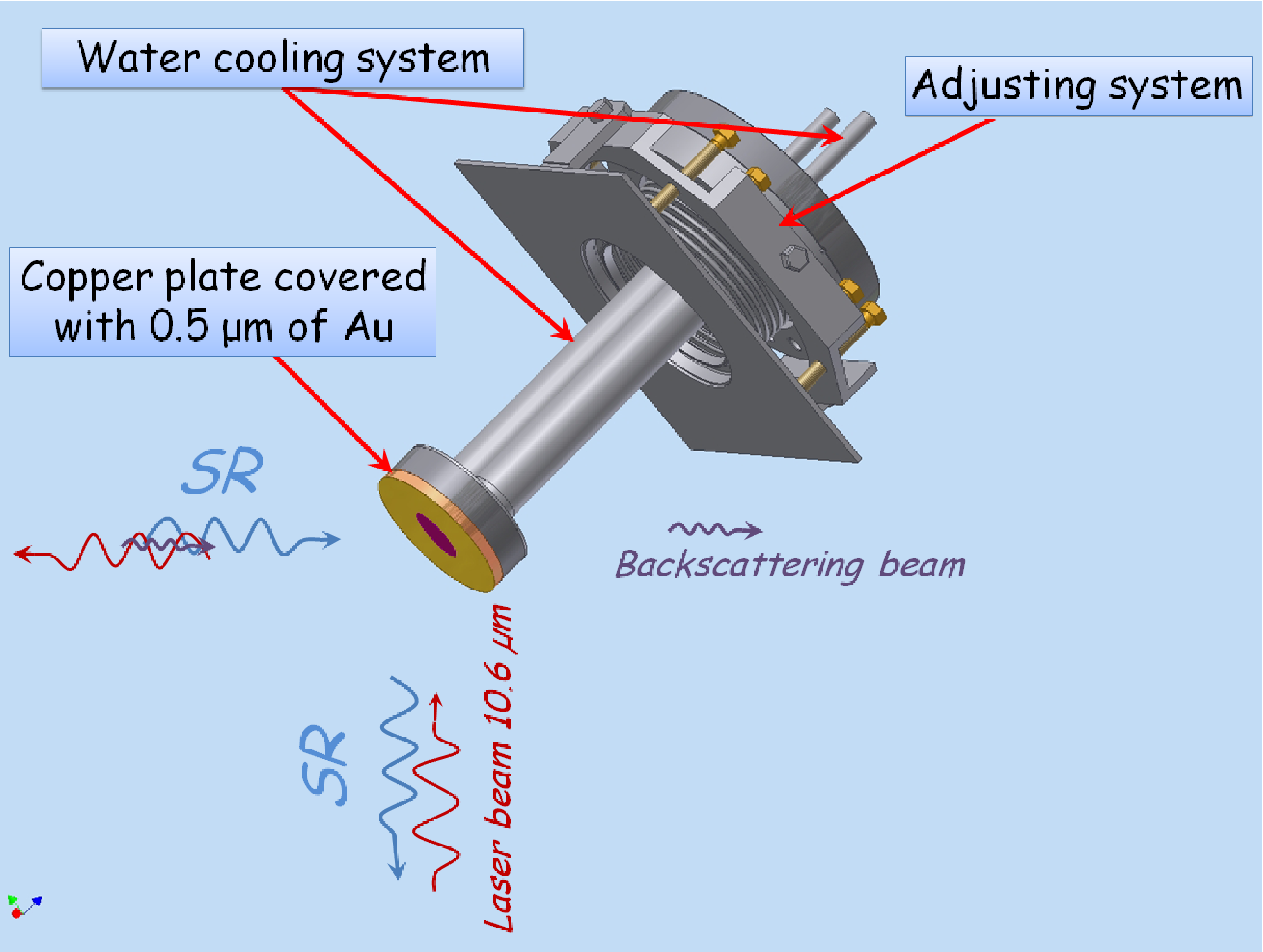}
\caption{Copper mirror.}
\label{mirror}
\end{figure}

The copper mirror design is shown in
Figure~\ref{mirror}. The mirror can be turned by bending the flexible bellows,
so the angle between the mirror and the laser can be adjusted as necessary. 
Note, the copper mirror protects the viewport against high 
power synchrotron radiation due to low reflectivity of high energy photons 
(less than 1\%) from a metallic surface. 
Synchrotron radiation (SR) photons heat the mirror. The extraction of heat is 
provided by a water cooling system. Cooling capacity is about 400 W. To 
prevent adsorption of residual gas molecules on the mirror surface, it is 
covered with a 0.5 $\mu$m thick gold layer.

\section{Conclusion}

The vacuum system for injection of laser beam in accelerator vacuum chamber
was designed. The system provides insertion of the light with wavelength in
the range from 0.45 to 20 $\mu$m. The system is used for calorimetric 
measurement of the VEPP-2000, VEPP-4M, BEPC-II colliders beams energy using 
CBS method. After installation of the system at colliders and backing out at 
$250^\circ C$  during 24 hours the pressure of about $10^{-10}$ Torr was 
obtained.

\acknowledgments

This work was supported by Russian Science Foundation 
(project N 14-50-00080).

\end{document}